\documentclass[a4paper,fleqn]{cas-sc}

\usepackage[authoryear]{natbib}
\usepackage{rotating}
\usepackage{lineno}
\usepackage{setspace}
\usepackage{appendix}
\usepackage[table]{xcolor}
\usepackage{cleveref}

\def\tsc#1{\csdef{#1}{\textsc{\lowercase{#1}}\xspace}}
\tsc{WGM}
\tsc{QE}
\tsc{EP}
\tsc{PMS}
\tsc{BEC}
\tsc{DE}

\begin{document}
\let\WriteBookmarks\relax
\def\floatpagepagefraction{1}
\def\textpagefraction{.001}

\shorttitle{K.I. Dale et al. 2026}

\shortauthors{K.I.Dale et al.}

\title [mode = title]{Oxidation Constraints on Terrestrial Planet Formation from a Ring}                      
\tnotemark[1]

\tnotetext[1]{This project was funded by the ERC HolyEarth, grant number: 101019380}

%
\author[1]{Katherine I. Dale\footnote{Now at: Bayerisches Geoinstitut, University of Bayreuth, Bayreuth, 95447, Germany}}[type=editor,
                        auid=000,bioid=1,
                        orcid=0000-0001-7511-2910]

\cormark[1]

\ead{katherine.dale@unibayreuth.de}


\credit{Conceptualisation of this study,
Methodology, code, manuscript and analysis of results}
\affiliation[1]{organization={Université Côte d'Azur, Observatoire de la Côte d'Azur, CNRS, Laboratoire Lagrange},
    city={Nice},
    postcode={06302}, 
    country={France}}

\cortext[cor1]{Corresponding author}

\author[3,1]{Alessandro Morbidelli}
\credit{Conceptualisation of this study, manuscript,
methodology and code }
\author[2]{Gabriel Nathan\footnote{Now at: Jet Propulsion Laboratory, California Institute of Technology, Pasadena, CA 91109, USA}}
\credit{Data and editorial}
\author[1]{Jason Woo}
\credit{Data}

\author[4]{David Nesvorný}
\credit{Data and editorial}

\author[5]{David C. Rubie}
\credit{Methodology and code}

\affiliation[2]{organization={Department of Earth \& Environmental Sciences, Michigan State University},
    city={East Lansing},
    postcode={MI 48824}, 
    country={USA}}

\affiliation[3]{organization={Collège de France, CNRS, PSL Univ., Sorbonne Univ.},
    city={Paris},
    postcode={75014}, 
    country={France}}

\affiliation[4]{organization={Solar System Science \& Exploration Division, Southwest Research Institute},
    city={Boulder},
    postcode={CO 80302}, 
    country={USA}}
    
\affiliation[5]{organization={Bayerisches Geoinstitut, University of Bayreuth},
    city={Bayreuth},
    postcode={95447}, 
    country={Germany}}

\begin{abstract}
The present-day solar system comprises meteorites with varying oxidation levels, derived from different parent bodies.  Previous studies (e.g. \citealp{Rubie2011}) of the partitioning of siderophile elements between mantle and core during planetary growth and differentiation showed that Earth must accrete reduced bodies first and oxidised bodies later. Here we show that, if the terrestrial planets formed from a narrow ring of planetesimals, this condition is not fulfilled, whatever heliocentric gradient of oxidation is assumed in the ring. The reason is that planetary embryos quickly accrete planetesimals from the whole width of the ring, incorporating both reduced and oxidised material. The partially oxidised state of all planetary embryos leads to mismatches with the composition of the bulk silicate Earth (BSE) because oxygen fugacity strongly affects the partitioning of siderophile elements. We demonstrate that reproducing the BSE composition requires reduced and oxidised reservoirs to remain segregated until embryo formation is almost complete. The delivery of oxidised material to the terrestrial planet-forming ring towards the end of the disc’s lifetime is therefore a key requirement of any successful dynamical model of terrestrial planet formation. 
\end{abstract}



\maketitle
\section{Introduction}
The present-day solar system comprises material of varying oxidation states, from highly reduced enstatite chondrites and aubrites to water-rich, highly oxidised carbonaceous chondrites such as CIs. Ordinary chondrites have an intermediate oxidation state. 
Previous studies of the chemical differentiation of a growing terrestrial planet showed that the composition of the bulk silicate Earth (BSE) can be reproduced only if our planet accreted reduced material during the initial 60-70\% of its growth and oxidised material later (e.g. \citealp{Wade2005,Rubie2011,Fischer2017,2021Icar..36514497M,Liu2023}). This is because the partitioning of some siderophile elements is strongly affected by oxygen fugacity. \citet{Siebert2013} claimed that Earth could form fully from oxidised material like ordinary or carbonaceous chondrites and still achieve the current oxidation state of the Earth’s mantle, but this is probably because the effects of Si partitioning into the core are not accounted for \citep{Rubie2015}. From the analysis of the I/Pu ratio, \citet{Liu2023} also concluded that Earth predominantly accreted reduced and volatile-poor differentiated planetesimals, followed by a secondary phase of accretion of volatile-rich undifferentiated bodies. However, the accretion of too much reduced material often results in an Earth's core that incorporates of too much Si (e.g. \citealp{Rubie2015}).

This temporal sequence in the accretion of reduced and oxidised material is achieved by the Grand Tack model \citep{Walsh2011}, where the radial migration of Jupiter depletes the asteroid belt by pushing material towards 1 AU, if one postulates an oxidation gradient with respect to heliocentric distance \citep{Rubie2015}. In this case, Earth starts accreting local, reduced material, and later, when Jupiter pushes planetesimals from the asteroid belt to the Earth’s region, it gets its share of oxidised material. This is the reason why the Grand Tack model was so successful from the geochemical point of view \citep{Rubie2015,Rubie2016,Jennings2021,Blanchard2022,Dale2023}.

From the dynamical point of view, however, the Grand Tack is no longer the favoured model, because the long-range outward migration of Jupiter appears unlikely, particularly in a low-viscosity disc \citep{Griveaud2024}. Moreover, the strong mixing induced by the inward migration of Jupiter tends to produce compositionally similar Earth and Mars, in contradiction with observations \citep{Woo2018}. Meanwhile, an improved understanding of planetesimal formation via the streaming instability \citep{Youdin2006} led to the realisation that the first planetesimals should have formed only at specific locations in the disc, i.e. in rings \citep{Drazkowska2016,Morbidelli2022}. Thus, the requirement that the building material had to be concentrated near 1 AU to reproduce the characteristic mass distribution of the terrestrial planets, with massive Earth and Venus at the centre and small Mercury and Mars on the wings \citep{Hansen2009}, does not seem to imply any specific migration pattern of Jupiter; it is rather a natural consequence of how planetesimals formed. The growth of terrestrial planets from a ring of material is now known as the “ring model” and has been investigated from the dynamical point of view in several publications \citep{Hansen2009, Nesvorny2021, Izidoro2022,Woo2023,Woo2024}. 

In \citet{Dale2025}, we have modelled the chemical composition of the BSE resulting from the ring model. Our analysis was based on N-body simulations of \citet{Nesvorny2021}, which began with a narrow ring of planetesimals and embryos, as well as a low mass extension of the planetesimal population towards the asteroid belt. This initial mass distribution was then split into four distinct compositional regions in order of increasing heliocentric distance: (i) a region of highly reduced, volatile-free objects, enriched in elements more refractory than Si with respect to solar composition, (ii) an enstatite chondrite region, (iii) an ordinary chondrite region and (iv) a CI chondrite region. While the existence of the innermost region is partially speculative, many studies have suggested the need for a missing reservoir of material in isotopic and compositional space (e.g. \citealp{Morbidelli2020,Burkhardt2021,Marrocchi2025}) that is only sampled within the Earth. In an attempt to account for such material, we included this innermost region. In our model, it was characterised by solar ratios of all elements except for volatiles, which are entirely depleted, and elements more refractory than Si, which are enriched to account for the Earth's supra-chondritic Al/Si and Mg/Si ratios \citep{Rubie2011,Dauphas2015,Morbidelli2020}. This material was assumed to form closer to the Sun to be more reduced than even enstatite chondrites (all Fe is in metal and 15-20\% of Si is metallic, see \autoref{tab:Oxidation}). While material this reduced is not something we see commonly in the meteorite record, if we accept the existence of a missing reservoir then this speculative material seems far less wild of a proposition. Using these compositions, we were able to reproduce the chemical composition of the BSE, provided the boundaries between these regions were such that at least one embryo accreted by Earth had an ordinary chondrite bulk composition. This is consistent with the requirement that Earth accreted mostly from reduced material, but with a non-negligible oxidised contribution in the later phases of its growth. 

In this work, we repeat this analysis on the simulations named 'shallow inner - Case 1 - 15 Myr' from \citet{Woo2023,Woo2024} (first line in Table 1 of \citealp{Woo2024}). We consider Earth analogues as planets with a mass within 10\% of that of the Earth by the end of the simulation, and thus many of the simulations were rejected for not meeting this criterion. These simulations are, in principle, superior to those of \citet{Nesvorny2021} because they feature the self-consistent growth of the embryos from the original planetesimal population, whereas \citeauthor{Nesvorny2021} assumed the existence of embryos in their initial conditions. For our purposes the difference may be significant: adopting radial zoning as in \citet{Dale2025}, using \citeauthor{Nesvorny2021} initial conditions means that embryos have a 'pure' composition (depending on the zone they are originally located in), whereas when using \citeauthor{Woo2023} initial conditions the embryos turn out to have a hybrid composition, resulting from the mixing of the planetesimal materials they form from. Thus, it is not obvious that the chemistry of the BSE can still be reproduced.

In \autoref{methods}, we start by recalling our planetary differentiation model from our previous works. In \autoref{results}, we present the results assuming radial zoning of the planetesimal ring of \citet{Woo2023,Woo2024}, showing that no zoning of the ring can lead to a satisfactory reproduction of the BSE chemistry. Then, in \autoref{LMR}, we discuss an alternative model where the planetesimals that participate in the construction of the embryos during the first 1.5 Myr are all reduced, whereas those incorporated later can have different degrees of oxidation. This new scenario leads to satisfactory results. A discussion of the mechanisms for this late delivery of oxidised material is included in \autoref{results}. The significance of this model is included in \autoref{Conc}, which summarises the conclusions. 

\section{Methods} \label{methods}
\subsection{The Ring Model Simulations} \label{ring}
\begin{figure}[hbt!] 
    \centering
    \includegraphics[width=0.5\linewidth]{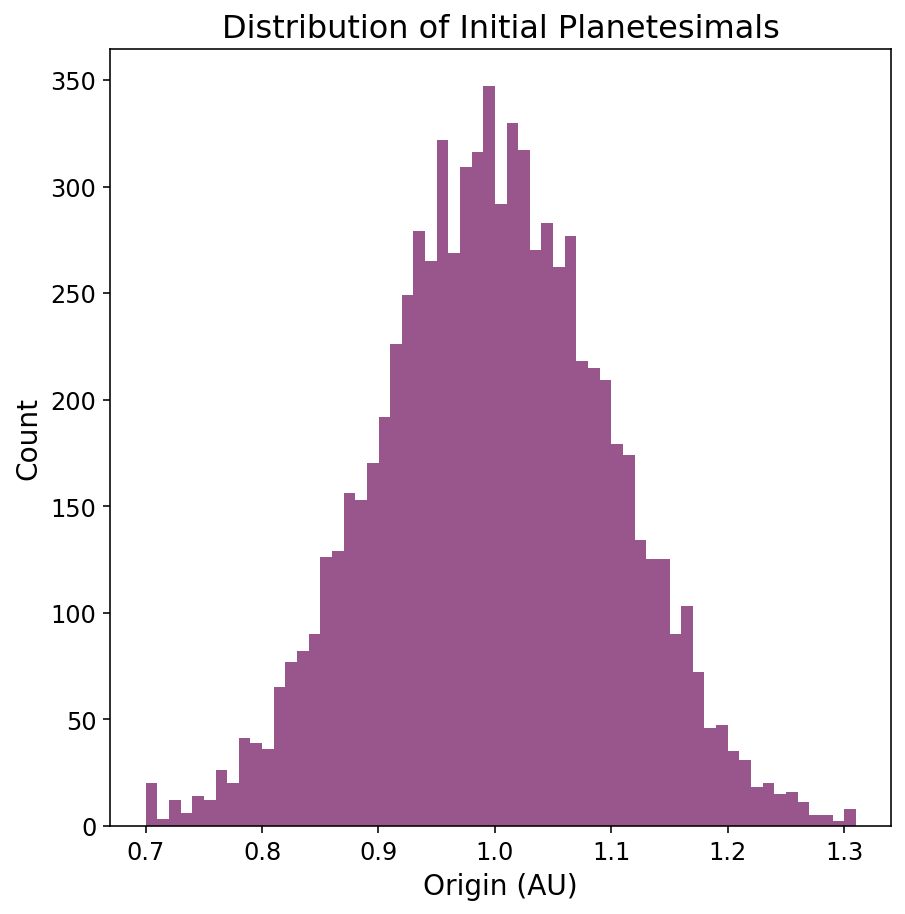}
    \caption{The initial distribution of the planetesimals of 'shallow inner - Case 1 - 15 Myr' from \citet{Woo2023,Woo2024}. There are five different initial distributions from 'shallow inner - Case 1 - 15 Myr' tested in our model but each is very similar: a bell-shaped curve around 1 AU extending between 0.7 and 1.3 AU.}
    \label{fig:Origin_Plots}
\end{figure}

This study uses N-body accretion scenarios coupled with a model of metal-silicate equilibration during the growth of terrestrial planets to test whether a certain sequence of collisions can recreate the BSE. We apply the same methods as \citet{Dale2025} to test the ring model simulations of terrestrial accretion presented in \citet{Woo2023,Woo2024}. These simulations start from a narrow, dense ring of planetesimals located between 0.7 and 1.3 AU. These initial distributions are shown in \autoref{fig:Origin_Plots}. The planetesimals begin to accrete to form embryos within the gas phase of the disc. This accretion continues after gas removal until a terrestrial planet system similar to that of the Solar System is formed. This is in contrast to the simulations of \citet{Nesvorny2021}, tested in \citet{Dale2025}, which began after the removal of the gas disc and thus assumed the existence of already-formed embryos in the ring as the initial conditions of the N-body simulations. Our method, therefore, had to be adapted slightly to account for this difference. Previously, embryos present in the initial conditions were differentiated at the beginning of the simulation, defining a core and mantle composition based on this differentiation, with fully oxidised bodies not forming a core. We now build these embryos from a collection of planetesimals at the beginning of the model. The composition of each individual planetesimal is added to the corresponding embryo for the initial time interval of the simulations, assigning each embryo a bulk composition that is cumulative of all the planetesimals it has accreted during that time. At the end of this time interval, each embryo is then differentiated starting from this bulk composition. This is intended to represent the continuous differentiation which occurs during the lifetime of $^{26}$Al and other heat-producing isotopes, leading to very hot planetary interiors. During this initial differentiation, the entire silicate mass is assumed to equilibrate with the entire metal mass, as typical in a single-stage differentiation scenario.

Each planetesimal can have one of four compositions depending on its initial semi-major axis. These regions are as described in the introduction, from  \citet{Dale2025}, in order of increasing heliocentric distance: (i) a region of highly reduced, volatile-free objects, enriched in elements more refractory than Si with respect to solar composition, (ii) an enstatite chondrite region, (iii) an ordinary chondrite region and (iv) a CI chondrite region. For full compositions of all regions, see \cref{tab:Compostions,tab:Oxidation}.
Relative contributions of each of the disc compositions are governed by free fitting parameters, which are determined by a least-squares minimisation routine designed to match the simulated Earth mantle to the BSE. This is achieved by optimising five fitting parameters, four of which relate to composition. Three of these parameters split the ring into the four compositional regions (D1, D2 and D3, such that D1 < D2 < D3). A fourth, denoted EF, determines the enrichment of the innermost region in elements more refractory than Si. By varying these four parameters, the model controls how much material from each compositional region the Earth accretes as well as the enrichment of elements more refractory than Si inherited from that innermost region. The values of these parameters are entirely in the control of the fitting algorithm within the model, in such a way that it can create an Earth from a variety of different compositional fractions, including an Earth potentially entirely made from one compositional type. A fifth fitting parameter, PEF, is also used following the investigations of \citet{Dale2025} and defines the depth at which planetesimals equilibrate. A description of this equilibration and the need for this fifth free fitting parameter is included in \autoref{MSE}. 

\subsection{Metal-Silicate Equilibration} \label{MSE}
The model of metal-silicate equilibration, which is used to model differentiation during planet formation, is based on that of \cite{Rubie2015} with the adaptations of \cite{Dale2023}. Within the model, there are two types of collisions: embryo-embryo (or giant impact) collisions and planetesimal-embryo collisions. We define a giant impact as one which melts >5\% of the target mantle on impact (as calculated using the scaling laws of \citealp{Nakajima2021}). 

Following an embryo-embryo collision, a magma pond is formed. The core of the impactor separates from the impactor mantle and descends through this magma pond. It equilibrates at the base of the magma pond with a localised portion of the silicate as defined by \cite{Deguen2014} (see figures in \citealp{Rubie2015,Rubie2025}). The amount of melt produced and the pressure at the base of the target's magma pond are defined by the scaling laws of \cite{Nakajima2021}, while the temperature is defined as halfway between the liquidus and solidus of peridotite corresponding to this pressure. The oxygen fugacity during equilibration is calculated via mass balance as described in \cite{Rubie2011} and evolves self-consistently by the partitioning of Fe between metal and silicate at the given conditions. Following equilibration, any metal that has not been oxidised is added to the target core. The magma pond then hydrostatically relaxes into a surface global magma ocean. This magma ocean is assumed to contain the silicates from the target involved in core equilibration, and a fraction of the unequilibrated silicates and of the impactor silicates. It also contains all the planetesimal material accreted to either impactor or target embryo between giant impacts. This is because, on impact, planetesimals do not cause significant melting and thus their material is stored at the surface of an embryo until a giant impact occurs. Then the metal they contain is equilibrated in the surface global magma ocean following the 'dispersed' metal-silicate fractionation model, as described in detail in \cite{Rubie2025}. The metal disperses into small droplets in the vigorously convecting magma ocean, where it equilibrates with the silicate magma ocean and segregates after being progressively swept into the mechanical boundary layer at the base of the magma ocean. Here, equilibration occurs progressively, with unoxidised metal added to the target core. 
 
 \begin{figure}[hbt!] 
    \centering
    \includegraphics[width=0.65\linewidth]{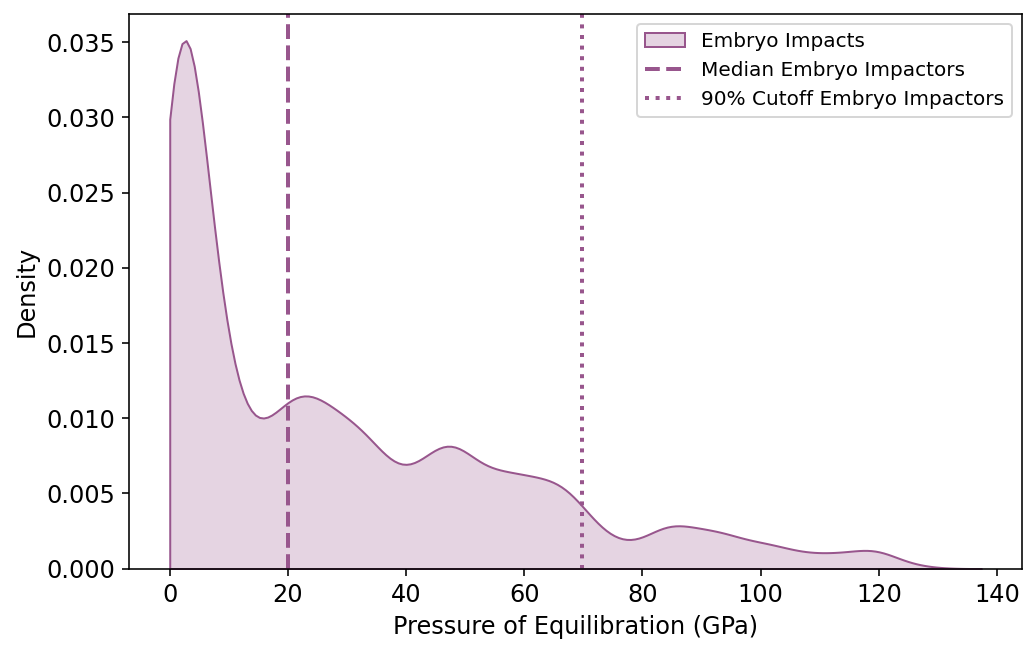}
    \caption{The pressure at which embryo cores equilibrate following giant impacts (impacts where both objects are larger than the standard planetesimal size). Most equilibration happens at 20 GPa or less (as most impacts are early when there are lots of smaller embryos in the protoplanetary disc). Only 10\% of embryo cores equilibrate at pressures >70 GPa (the point at which our partitioning laws become less reliable), which equates to less than one equilibration event per final planet with mass $\sim$ 1 Earth mass.}
    \label{fig:PressureDis}
\end{figure}

 However, the magma ocean crystallises rapidly (likely within 5 Myr \citealp{ElkinsTanton2008}), so that a significant portion of the metal delivered by planetesimals between giant impacts may equilibrate at lower pressures and temperatures than those corresponding to the initial magma ocean depth. While the magma ocean is crystallising, its base is rising towards the surface. Thus, the average pressure at which planetesimal material equilibrates will not be equal to that at the base of the initial surface global magma ocean but will be somewhat smaller, reduced by a factor dubbed PEF (for planetesimal equilibration factor). In the absence of a realistic model of magma ocean crystallisation, we take PEF to be the fifth (and last) fitting parameter in our model and ranges between 0 and 1.

Whether a collision is embryo-embryo or embryo-planetesimal, metal-silicate equilibration is governed by partitioning laws calculated from thermodynamics and experiments. In order to be able to compare our results with those of our previous studies directly, we use the metal-silicate partition coefficient parametrisations of \citep{Rubie2015}. For some siderophile elements, these parametrisations are based on experimental results performed up to 25 GPa \citep{Mann2009}. While this is considerably lower than the maximum pressures of metal-silicate equilibration in the present study, it is representative of the pressures at which most equilibration occurs (see \autoref{fig:PressureDis}). However, large extrapolations for higher pressures are not necessary in the case of the most important elements in this study, O and Si. The partitioning of O is determined using the model of Frost et al. (2010), which is based on a thermodynamic model, the results of which are shown to be consistent with O partitioning determined in multianvil and diamond anvil cell (DAC) experiments up to at least 70 GPa and 3500 K. Recent work has shown that these results are still in line with experimental conclusions (Frost et al. (in review)). The partitioning of Si is from \citet{Mann2009} and is based on fitting the results of multianvil experiments (obtained up to 25 GPa) with previously published DAC results obtained at 55 and 93-97 GPa. For some siderophile elements, extrapolations of low-pressure data are consistent with high-pressure results (e.g. Ni). For W and Mo partitioning we use \citet{Jennings2021}. While their data may not be valid up to the high pressures of the DAC results of \citet{Huang2021}, their data (when extrapolated) agrees well with that of \citet{Huang2021}. Additionally, they account the effect of C concentration, something that is very important for Mo and W partitioning. Notice that, in our results, we will find misfits in major element compositions such as SiO$_2$ in the BSE. Regardless of the partitioning we use for W and Mo, this misfit would still remain. Therefore, for consistency with previous works (e.g. \citet{Dale2023,Dale2025}), we use the extrapolation of the \citeauthor{Jennings2021} results.

The calibrations of partitioning at different pressures and temperatures used in this model are just some of the possible calibrations that could have been used in this study. While we use the parametrisations listed above to allow comparison between this work and previous work we have also explored what effect other parametrisations may have on our results. In order to test the influence of multi-anvil (MA) compared to diamond-anvil cell (DAC) calibrations on core-mantle partitioning, we compare our parametrisations to those of \citet{Sossi2025}. The later describes the partitioning of Si, Fe, Ni and O based on the studies of \citet{Ricolleau2011,Rubie2011,Siebert2013} and \citet{Fischer2015}.

For simplicity in our comparisons, we test these different metal-silicate partitioning laws on a single-stage core-formation model with homogenous accretion. For a homogeneous accretion, the material forming the Earth has to have the bulk Earth composition. Thus, we begin with an undifferentiated body with a composition of the bulk Earth, as given in \citet{Fischer2025}, for Si, Al, Ca, Mg, Fe, Ni, and O. These elements comprise approximately 98.5 wt\% of the Earth.  We compute the SiO$_2$ and FeO concentrations in the mantle after equilibration at a given pressure and at a temperature corresponding to the peridotite liquidus at that pressure \citep{Andrault2011}, for direct comparison with \citet{Sossi2025}. It should be noted that this temperature has a slight difference compared with the one used throughout the rest of this work, where the temperature lies halfway between the solidus and liquidus of peridotite at a given pressure following \citet{Rubie2015}. Additionally, while the \citeauthor{Mann2009} law has a pressure dependency, the one used by \citeauthor{Sossi2025} does not. As is traditional in the single-stage differentiation scenario, we assume full equilibration between metal and silicate.
 
 \begin{figure}[hbt!] 
    \centering
    \includegraphics[width=0.8\linewidth]{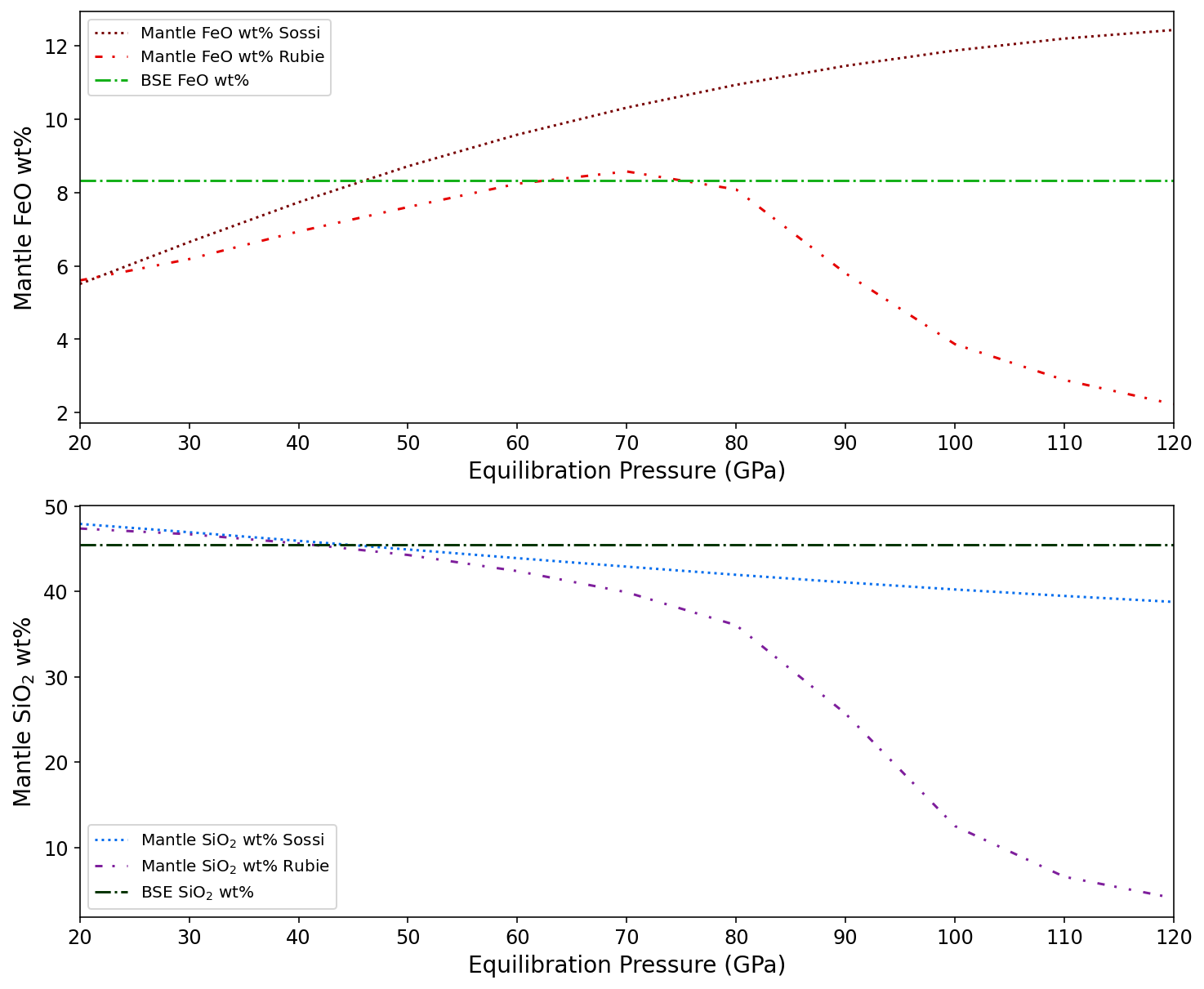}
    \caption{Variation in final Earth mantle wt\% of FeO (top) and SiO$_2$ (bottom) as a function of pressure from 20 GPa to 120 GPa with two different partitioning laws across two scenarios. All bodies have bulk Earth composition. Dotted lines show the mantle concentrations using the law of \citet{Sossi2025} for single-stage equilibration. Dot-dot-dashed lines show single-stage equilibration following the partitioning laws used in \citet{Rubie2011,Rubie2015} and the main text of this work. The green dot-dashed lines on both figures show the BSE composition from \citet{Fischer2025}.}
    \label{fig:PressurePlots}
\end{figure}

The results are illustrated in \autoref{fig:PressurePlots}. The dotted curves (in red and blue, respectively) show the FeO and SiO$_2$ concentrations given by the partitioning law presented in \citet{Sossi2025}, while the dot-dot-dashed curves depict the values obtained applying the partitioning laws of \citet{Mann2009} used in \citet{Rubie2011,Rubie2015}. The remaining dot-dashed horizontal lines show the BSE composition of \citet{Fischer2025}. While the model of \citet{Sossi2025} shows a higher FeO than that of the \citet{Rubie2015} model, up to $\sim$70 GPa the curves show the same trend with the FeO concentration increasing and the SiO$_2$ concentration decreasing with increasing pressure. For P>70 GPa, the FeO concentration starts to decrease according to the partition law in \citet{Rubie2011}, while it keeps increasing according to the partition law in \citet{Sossi2025}.
 
However, equilibration events at pressures greater than 70 GPa do not play a large role in the formation of planets in the simulations of \citet{Woo2023,Woo2024}. \autoref{fig:PressureDis} shows the distribution pressures in of equilibration events across all forming planets (not just Earth) for giant impacts.  50$\%$ of all material equilibrates at 20 GPa or less (a pressure at which \citet{Rubie2015} and \citet{Sossi2025} parametrisations are almost identical in \autoref{fig:PressurePlots}). 90$\%$ of all equilibration events for embryos across all forming planets occur at $\leq$ 70 GPa. While it is true that Earth-like planets are more likely to experience these events, because larger pressures require larger objects, an Earth or Venus will experience, on average, $\leq$1 embryo equilibration event at > 70 GPa. While, in future work, other partitioning laws will be tested and used, the sharp turnover of the FeO and SiO$_2$ mantle concentrations at pressures > 70 GPa in the formulations of \citet{Rubie2015} which are used in this paper are unlikely to have a major impact on our final results. Further analysis of these parametrisations, including a discussion of the partial metal-silicate equilibration in multi-stage core-formation scenarios is included in \autoref{Homovhetero}.

\section{Results} \label{results}
\begin{figure} [hbt!] 
    \centering
    \includegraphics[width=1\linewidth]{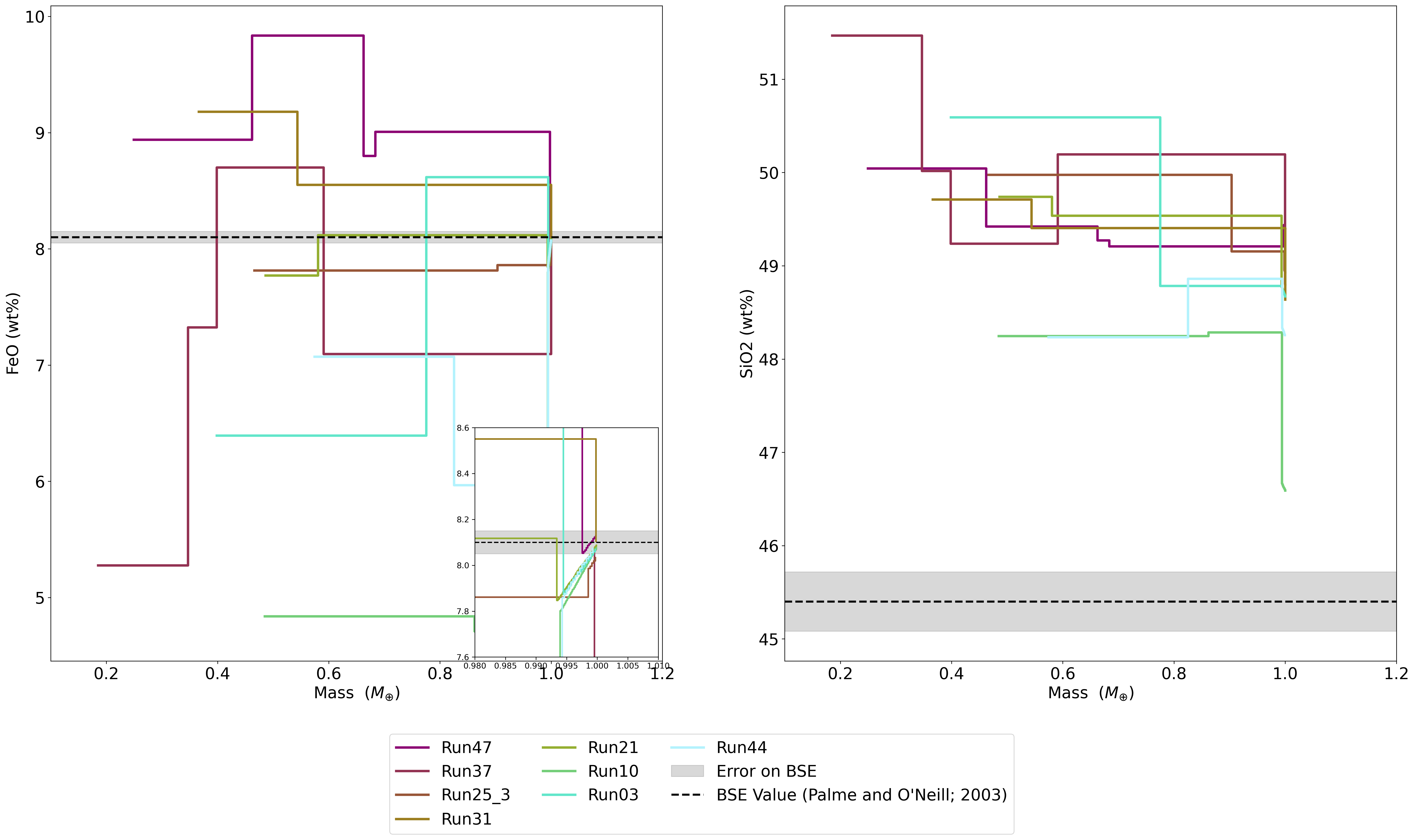}
    \caption{The changing mantle FeO and SiO$_2$ of 8 Earth analogues as they grow through the simulation. The black dashed line and the grey band show the BSE concentrations of FeO and SiO$_2$ \citep{Palme&ONeill2003} and the uncertainties on these values. The left-hand figure contains an insert that allows for a clearer view of the point at which the embryo masses reach 1 Earth Mass.}
    \label{fig:SiO2_FeO_IMR}
\end{figure}

36 simulations with the starting conditions described in \autoref{ring} were tested using the model described throughout \autoref{methods}. We assume that the phase of initial differentiation lasts 10 Myr. Although we fit for 12 different elemental oxides, the discussion will focus on FeO and SiO$_2$ concentrations in the final Earth-analogue mantles. The results for 8 of these simulations are shown in \autoref{fig:SiO2_FeO_IMR}. Simulations from these \citeauthor{Woo2023} ring models fit the FeO concentration of the BSE well across most of the simulations. SiO$_2$, however, is consistently too high at the end of the simulations. SiO$_2$ starts high in all simulations following initial differentiation and is changed, with a general downward trend, by subsequent equilibration events, but it never reaches the BSE value. This suggests that either our simulations have too few giant impacts or begin with too high SiO$_2$ content. The number of giant impacts is an outcome of the the simulation itself, and this problem exists across 35 of the 36 simulations (the remaining one fails to fit BSE FeO), with a variety of impact sequences. Thus, it seems unlikely that the low number of giant impacts is the cause of the mismatch between Earth-analogue mantles and the BSE. Si is removed to the core during giant impacts, however its siderophile nature increases as pressure and temperature increases for all partitioning laws we tested (see \autoref{fig:PressurePlots}). The high pressures (and thus temperatures) needed to move accreted Si to the core happen in very few impacts (see \autoref{MSE}) so are unable to sufficiently reduce the SiO$_2$ content in the mantle. A more reduced composition would move more Si to the core. This could perhaps be achieved by a greater quantity of our innermost refractory enriched material. However, our fitting code does not offer this possibility, suggesting that the overall fit to the BSE would degrade for other species, particularly FeO. The "best results" the the code converges to have a too high mantle SiO$_2$ concentration.

This result at first appears puzzling. In fact, \citet{Dale2025} showed that ring model simulations could successfully match the composition of the BSE, but these results seem to suggest the opposite. There is, however, a key difference between the simulations from \citet{Nesvorny2021} used in that paper and the simulations from \citet{Woo2023,Woo2024} used here that has an effect on the initial composition of embryos. The \citeauthor{Nesvorny2021} simulations begin with a ring of planetesimals and embryos, meaning that the large bodies had already formed. This resulted in each embryo having a 'pure' composition when the model splits the ring into four compositional regions. That is, each embryo is assigned the composition of only one of the four regions. This was usually the composition of either the refractory-enriched or enstatite chondrite region, with only one or two Earth-accreted embryos start having a purely ordinary chondrite-like composition. In contrast to this, embryos in the \citeauthor{Woo2023} simulations form only from planetesimals. Thus, the embryos can form by accreting planetesimals from the entire ring, acquiring a mixture of  all four compositions contained within it if the dynamics of the N-body simulations allow. 

\begin{figure}[hbt!] 
    \centering
    \includegraphics[width=0.7\linewidth]{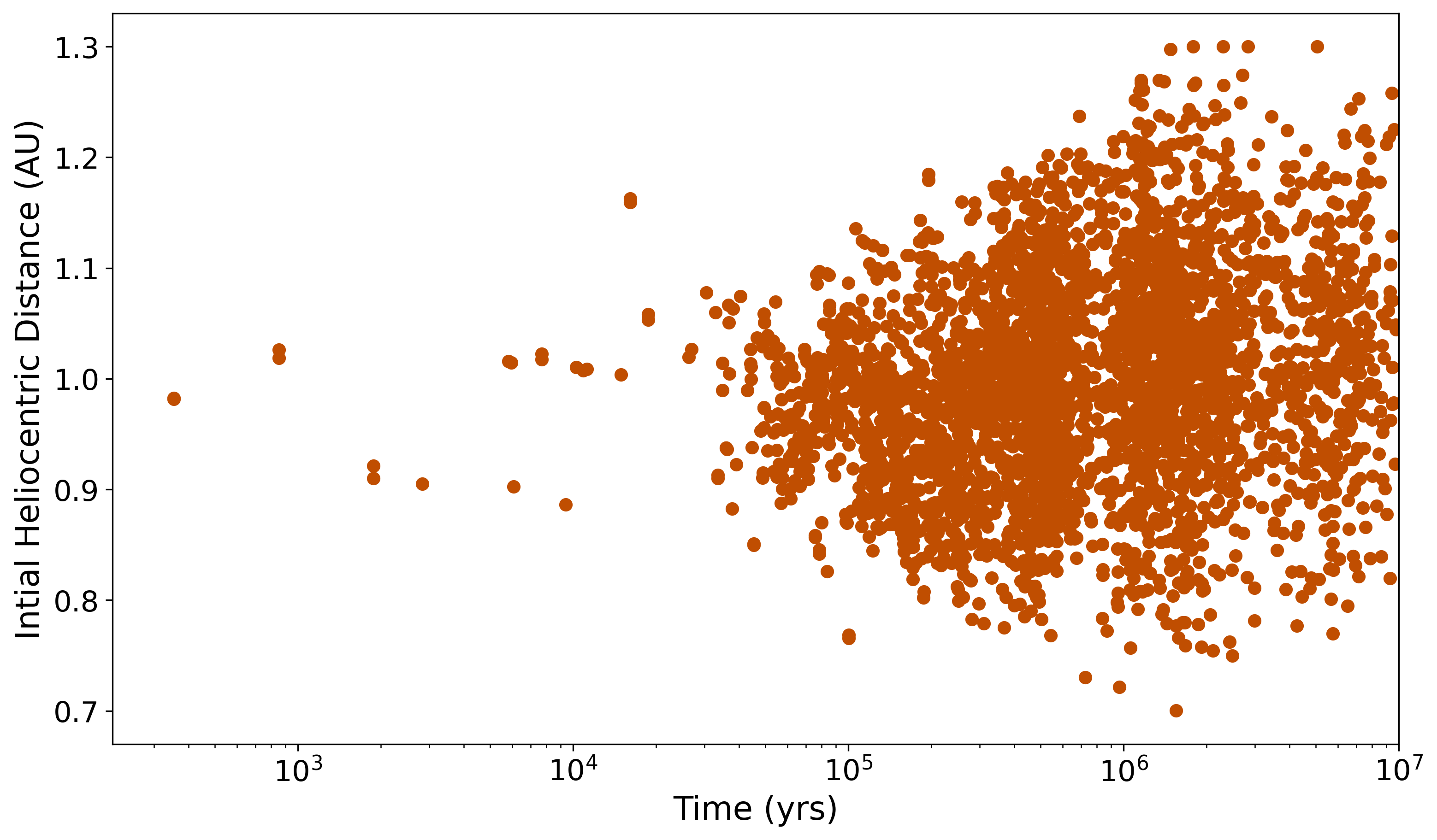}
    \caption{The material accreted in the first 10 Myr by an Earth-forming embryo in a typical ring model simulation. Material accretes from the whole width of the ring (0.7 - 1.3 AU) within the first 1.5 Myr of accretion.}
    \label{fig:Distances}
\end{figure}

This indeed happens. Within the first 10 Myr, the time when the gas is removed from the disc, the main Earth-forming embryo has accreted material from the entire width of the disc (\autoref{fig:Distances}) and thus, if Earth is to accrete any oxidised material, it has done so by this point. Reducing the duration of continuous differentiation to, for example, 1.5 Myr would not alleviate this problem. Indeed, \autoref{fig:Distances} shows that a typical embryo forming in the ring accretes material from as far as 1.2 AU within 200,000 yrs and material from the whole ring width by 1.2 Myr. This means that the Earth has to either be formed of only one material, and thus the whole ring is homogeneous, or be formed from a mixture of both reduced and oxidised material from the beginning. The model of metal-silicate equilibration prefers a mixture of oxidised and reduced material to form the Earth, and thus a heterogeneous ring. This makes all embryos already partially oxidised (rather than fully reduced as most of those in \citealp{Dale2025}) at the end of the continuous differentiation phase, and so are too rich in SiO$_2$ from the beginning. Having partially oxidized embryos violates the conclusions of \cite{Wade2005,Rubie2011,Fischer2017,2021Icar..36514497M,Liu2023}, that the Earth should have accreted first reduced material and later oxidised material. It may be suggested that, under the right conditions and with the right partitioning formulations, Earth could be formed from a homogeneous reservoir of partially oxidised material. This is explored in \autoref{Homovhetero} where we show that, due to the nature of multi-stage core formation, whatever partitioning law is used, the SiO$_2$ and FeO concentrations of the BSE can not be matched simultaneously starting from a single bulk Earth composition. This is because, if the whole mantle silicate is not available to equilibrate with the descending metal core (i.e. $k_{mantle} \neq 1$), not enough Si can be moved to the core. While the value of k is uncertain, most experiments put it at <0.5 except for very energetic impacts (see \citet{Landeau2021,Dale2023,ROHLEN2025}). This means we need a ring of reduced material to start with and a mechanism that can allow for oxidised material to be brought into the ring after the Earth has begun to accrete. 

A first possibility is that the oxidation state of the protoplanetary disc changed over time, while planetesimals continuously formed at the ring location. In this case, the first planetesimals could be reduced, and the late planetesimals could be oxidised. The meteorite record, however, does not support this hypothesis. There is no apparent correlation between formation time and oxidation state of the meteorite parent bodies \citep{Grewal2023}. In fact, as stated in the introduction, enstatite chondrites and aubrites are the most reduced meteorites, although they formed at radically different times (aubrites are achondrites and so their parent body formed much earlier). Similarly, many iron meteorites, debris of the cores of early-formed and differentiated planetesimals, reveal parent body oxygen fugacities comparable to those of ordinary chondrites \citep{Grewal2023}.

A second, more realistic possibility, is that the dynamical evolution brought material into the ring from a different, more oxidized region. \citet{Nesvorny2025} postulated the existence of two rings: one centred at 0.65 AU and one centred at 1.6 AU. The former could be made of reduced material, the second of more oxidized material. They showed that Earth started forming from the inner ring, together with Venus and Mercury, but then accretes about 30\% of its mass from the outer ring, which is the dominant one from which Mars formed. More recently, \citet{Goldberg2026} showed that the outer ring at 1.6 AU can be formed from the accumulation of planetesimals, initially formed throughout the asteroid belt (therefore with an oxidation state similar to that of ordinary chondrites), due to secular-resonance sweeping  the belt as gas is progressively reduced. Note that the delivery of planetesimals initially from the asteroid belt to the Earth forming region was also a characteristic of the Grand Tack scenario, which explains its success for geochemical simulations \citep{Rubie2015}, but that dynamical mechanism is no-longer considered valid. The scattering of carbonaceous planetesimals from the giant planet region to the terrestrial forming region \citep{Raymond2017}, cannot be a sufficient source of oxidized material, because isotopic constraints limit the mass accreted by the Earth from this source to be < 5\% \citep{Steller2022,Sossi2026}. Thus, as of today the models of \citet{Nesvorny2025} and \citet{Goldberg2026} are the most realistic to explain the delayed delivery of oxidized material to the Earth-forming region.

\subsection{Late Introduction of Oxidised Material} \label{LMR}
Here, to remain agnostic to the exact dynamical process for the late delivery of this oxidised material, we still use \citeauthor{Woo2024} simulations. In order to test whether this late addition of oxidised material can result in a match for the BSE mantle, we posit that the ring remained highly reduced for the first 1.5 Myr of planet formation. We assign all planetesimals that have a collision before 1.5 Myr the composition of the refractory enriched region, which means that all embryos are initially formed with a reduced composition. We then, allow all other planetesimals accreted to Earth later to have a composition from one of the four compositional regions. This means that all initial material, which is accreted in the first 1.5 Myr, is reduced, but those accreted afterwards can have any one of the four compositions and thus be more oxidised (i.e. from an OC- or CC-like reservoir) if that is what is required to fit the BSE composition. This mimics the delivery of material in \citet{Nesvorny2025,Goldberg2026}.

\begin{figure} [hbt!] 
    \centering
    \includegraphics[width=1\linewidth]{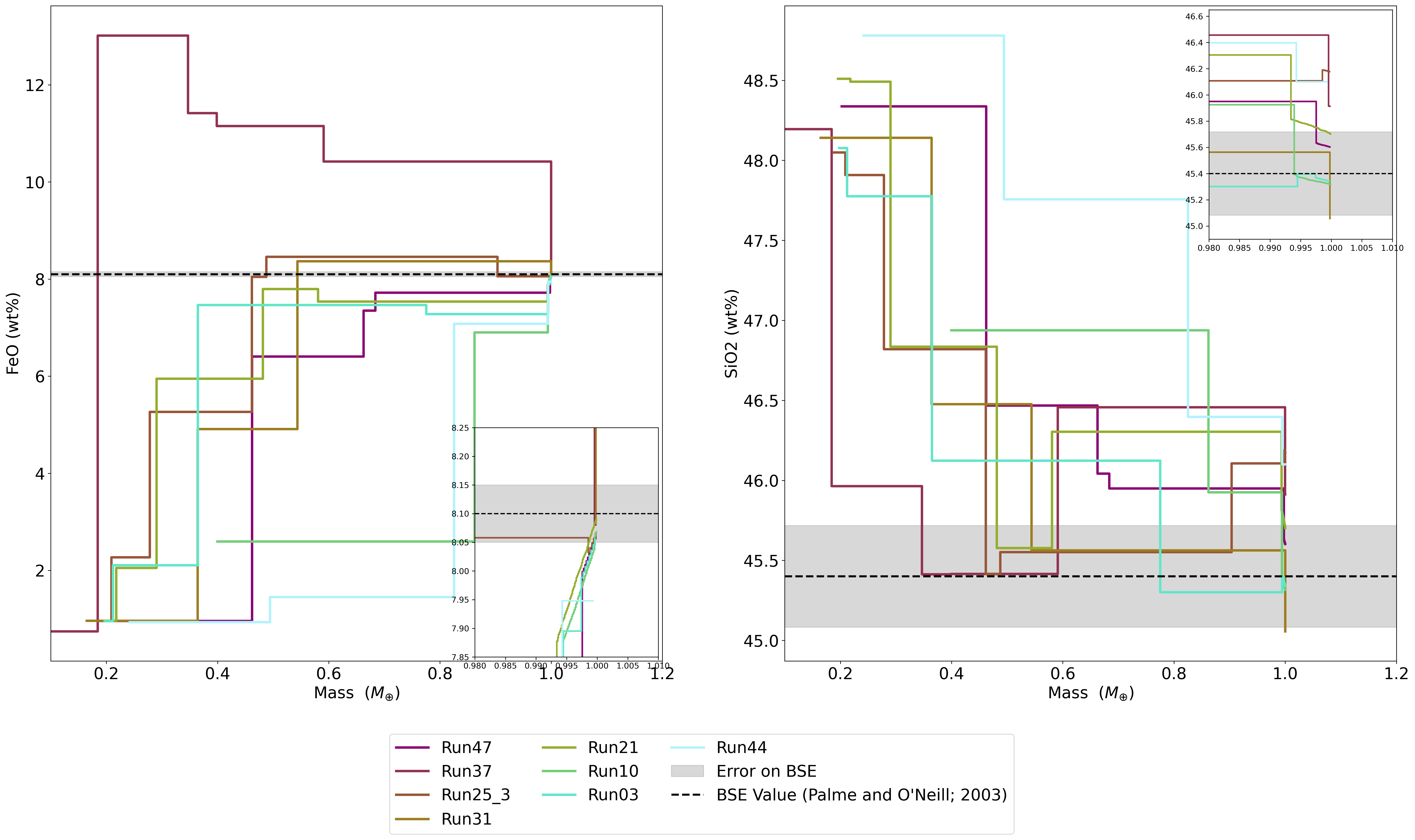}
    \caption{As \autoref{fig:SiO2_FeO_IMR} but for the simulations of \autoref{LMR} where the planetary embryos are all initially fully reduced and planetesimals accreted after 10 Myr can be both reduced or oxidised (see \cref{tab:Compostions,tab:Oxidation}). Both figures have an insert highlighting the point at which the proto-Earth is fully grown.}
    \label{fig:SiO2_FeO_LMR}
\end{figure}

We find that this results in a vast improvement in the ability of these ring model simulations to reproduce the BSE composition. The results for SiO$_2$ and FeO are shown in \autoref{fig:SiO2_FeO_LMR}. These show that not only can the BSE's FeO concentration be reproduced, but so can the SiO$_2$ concentration. This is because the initial compositions of all embryos are more SiO$_2$ poor than in the case previously discussed, which depicted more oxidised embryos. Not only does Earth start with less SiO$_2$ in its mantle, but the embryos it collides with are also less oxidised, leading to a quick reduction in overall mantle SiO$_2$. In addition to this, all 12 of the BSE elemental components can be fit in Earth analogue mantles in 5 of the simulations (see \autoref{fig:Sup4}). 

\begin{figure}[hbt!] 
    \centering
    \includegraphics[width=0.8\linewidth]{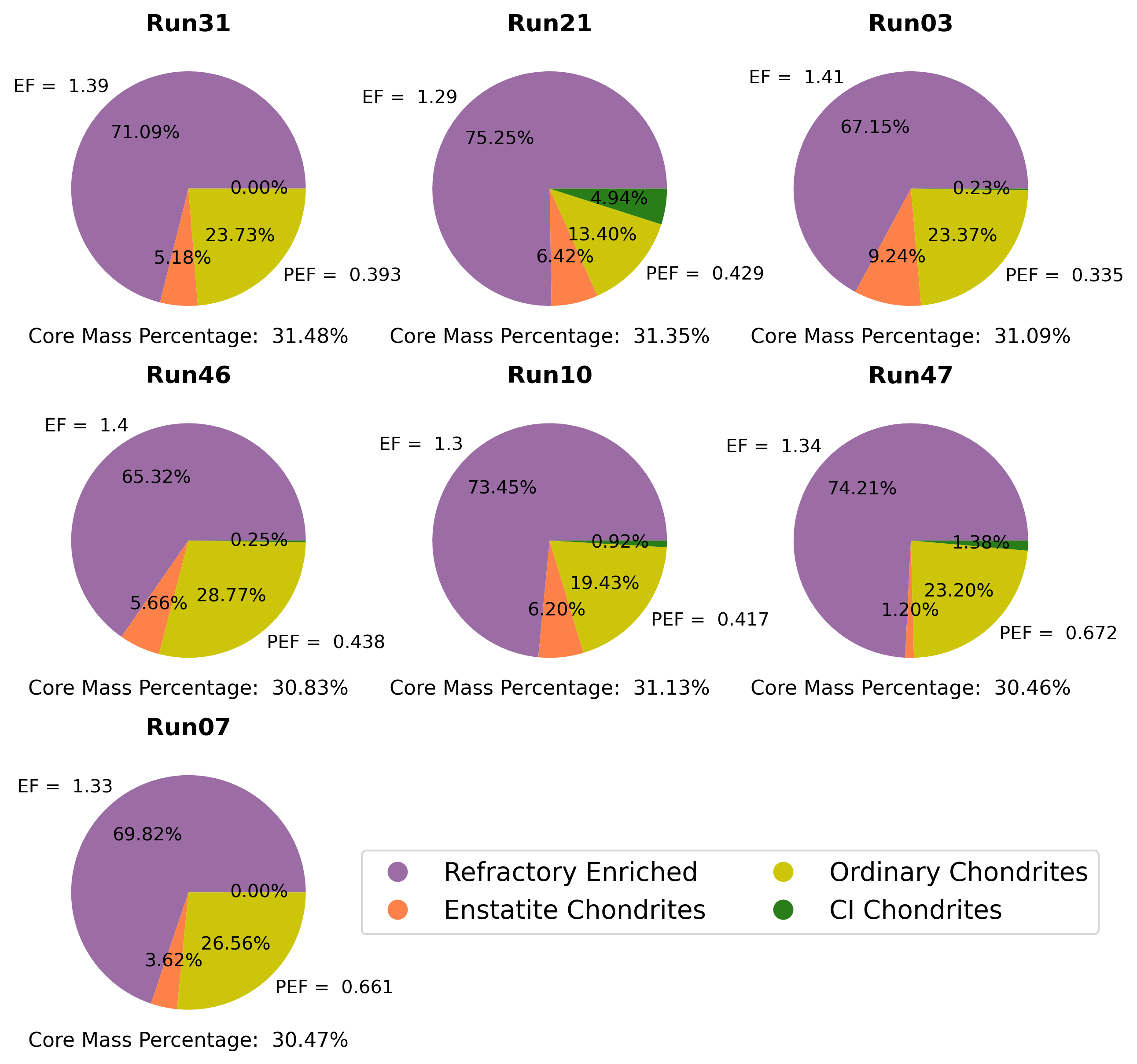}
    \caption{The contributions from the four compositional regions required to create the best possible Earth-analogue mantles formed by the late addition of oxidised material (OC- and CC-like material) to a reduced ring in a given simulation. Refractory enrichment (EF) of the innermost region is shown in the top left-hand corner of each chart, while the planetesimal equilibration factor (PEF) is shown on the lower right. Core mass, as a percentage of the final planet mass (1 Earth mass), is shown below each plot. }
    \label{fig:Sup1}
\end{figure}

 In general, the Earth-analogues produced by this model are made from 70-75\% material which is enriched by 30-45\% in elements more refractory than Si (see \autoref{fig:Sup1}). This further consolidates the conclusions of \citet{Dale2025} that some 'missing' material was an essential component for Earth's formation. This conclusion is derived only from chemical considerations, but is also the outcome of the analysis of the Earth's isotopic composition in the context of the solar system dichotomy \citep{Burkhardt2021}. Indeed, both isotopic and chemical evidence suggest that Earth does not look like any of the meteorite types we have in our collection suggesting we are 'missing' at least one component of the Earth-forming reservoir. In this work we find that such a missing component should have comprised a highly reduced material. The remaining accreted material comes largely from the enstatite and ordinary chondrite region, with 0-5\% accreting from CI chondrite material (\autoref{fig:Sup2}). In all simulations, the addition of oxidised material is dominated by OC-like material. As this missing component forms a large portion of the material accreted to Earth, it could be suggested that Earth formed entirely from it \citep{Drake2002}, provided that component had the composition of the bulk Earth. We test this possibility in \autoref{Homovhetero}, finding that this is possible but only if each equilibration event leads to full equilibration between metal and silicate ($k_{mantle}\sim 1$).

Notice that, had we used \citet{Sossi2025} partitioning law instead of \citet{Rubie2015}., the oxidised component should have been introduced a later in the simulation. This is because the former reduces the SiO$_2$ content of the mantle less during high pressure impacts, thus a longer period of highly reducing conditions is important in order to reduce SiO$_2$ concentration in an Earth-analogue's mantle.
 
\section{Conclusion} \label{Conc}
In this work, we have analysed the chemical composition of the BSE resulting from the ring model, using the simulations from \citet{Woo2023, Woo2024}, which start from a ring of planetesimals and feature the self-consistent growth of planetary embryos from mutual planetesimal collisions. As in \citet{Dale2025}, we have assumed radial compositional zoning through the ring. In the innermost zone, planetesimals are highly reduced, volatile-free and enriched relative to solar composition in elements more refractory than Si. In the other zones, planetesimals have the bulk composition of enstatite chondrites, ordinary chondrites and CI chondrites, in order of increasing heliocentric distance.

We show that, irrespective of the locations of the boundaries between these zones, the resulting chemical composition of the Earth analogue’s mantle is very different from the BSE composition. The reason is that all planetary embryos accrete planetesimals almost uniformly throughout the ring. Thus, they do not have significantly different compositions. They are either all fully reduced (if the entire ring is included in the two innermost zones) or all partially oxidised if a portion of the ring falls in the ordinary chondrite or CI zones. In either case, there is no switch during Earth’s accretion from highly reduced material to more oxidised material. Thus, in agreement with \citet{Rubie2011,Fischer2017,2021Icar..36514497M,Liu2023} and others, the BSE composition cannot be reproduced. 

Our solution to the problem proposes that all planetesimals participating in the growth of embryos during the first 1.5 Myr are highly reduced and refractory-enriched (as the bodies in the innermost zone in the previous case). The planetesimals that are accreted after 1.5 Myr can be either reduced or oxidised, depending on their initial heliocentric distance (following the zoning model defined above, with zone boundaries set as free fitting parameters). In this case, the initial growth of Earth is obviously from fully reduced material, while some oxidised material is accreted only later. This model allows us to reproduce satisfactorily the BSE composition. 

Of course, this secondary model is ad hoc and has been introduced for illustrative purposes only. However, recent models of terrestrial planet formation envision that the Earth's forming ring merged after a few My with a second ring of material of objects formed farther from the Sun \citep{Nesvorny2025,Goldberg2026}, support our two-stage compositional accretion scheme.

  Our study demonstrates that the delivery of oxidised material, consisting of material more oxidised than EC (OC- and CC-like) to the terrestrial planet-forming ring towards the end of the disc’s lifetime is a key requirement of any successful dynamical model of terrestrial planet formation. This shows the power of a holistic approach to terrestrial planet formation, combining dynamical and geochemical modelling and related constraints.

\section{Acknowledgements}
The authors would like to thank the reviewers and editors for their valuable feedback, which greatly improved this work. They also thank those who contributed through fruitful discussions, particularly Dan Frost for his input on partitioning laws.

Additionally, Katherine Dale acknowledges the support of the ERC Geoastronomy project, whose funding has enables her to continue her research.
\section{Bibliography}
\bibliographystyle{cas-model2-names}

\bibliography{references}
\newpage
\appendix
\renewcommand{\thefigure}{A.\arabic{figure}} 
\renewcommand{\theequation}{A.\arabic{equation}}
\renewcommand{\thetable}{A.\arabic{table}}
\renewcommand{\thesection}{A.\arabic{section}}
\setcounter{figure}{0} 
\setcounter{section}{0} 

\section*{Appendix}
\section{Homogenous vs heterogenous accretion scenario} \label{Homovhetero}
The need for heterogeneous accretion to achieve the correct chemical composition of the BSE was highlighted in \citet{Rubie2011}. That work showed that, in a multi-stage accretion scenario, it is not possible to reproduce the SiO$_2$ and FeO concentrations of the BSE if all the planetary embryos that build the Earth have the same bulk composition. 

It may be questioned whether this negative result is due to the specific partition laws of elements between silicate and metal used in that work (and in this work) or is a robust conclusion. 

We have shown in the main text that the partitioning of Si and Fe between metal and silicate are similar in \citet{Sossi2025,Rubie2015}. In particular, we notice that, with the partition law of \citet{Sossi2025}, the BSE composition can be reproduced from bulk-Earth composition, in a single stage differentiation scenario, if metal-silicate equilibration occurred at P=45 GPa, while no value of pressure allows reproducing both the FeO and the SiO$_2$ BSE concentrations if the law of \citeauthor{Rubie2011} is adopted. This seems to suggest that the need for heterogeneous accretion in \citet{Rubie2011} and other studies is just a consequence of the specific partition law that is used.

\begin{figure}[hbt!] 
    \centering
    \includegraphics[width=0.8\linewidth]{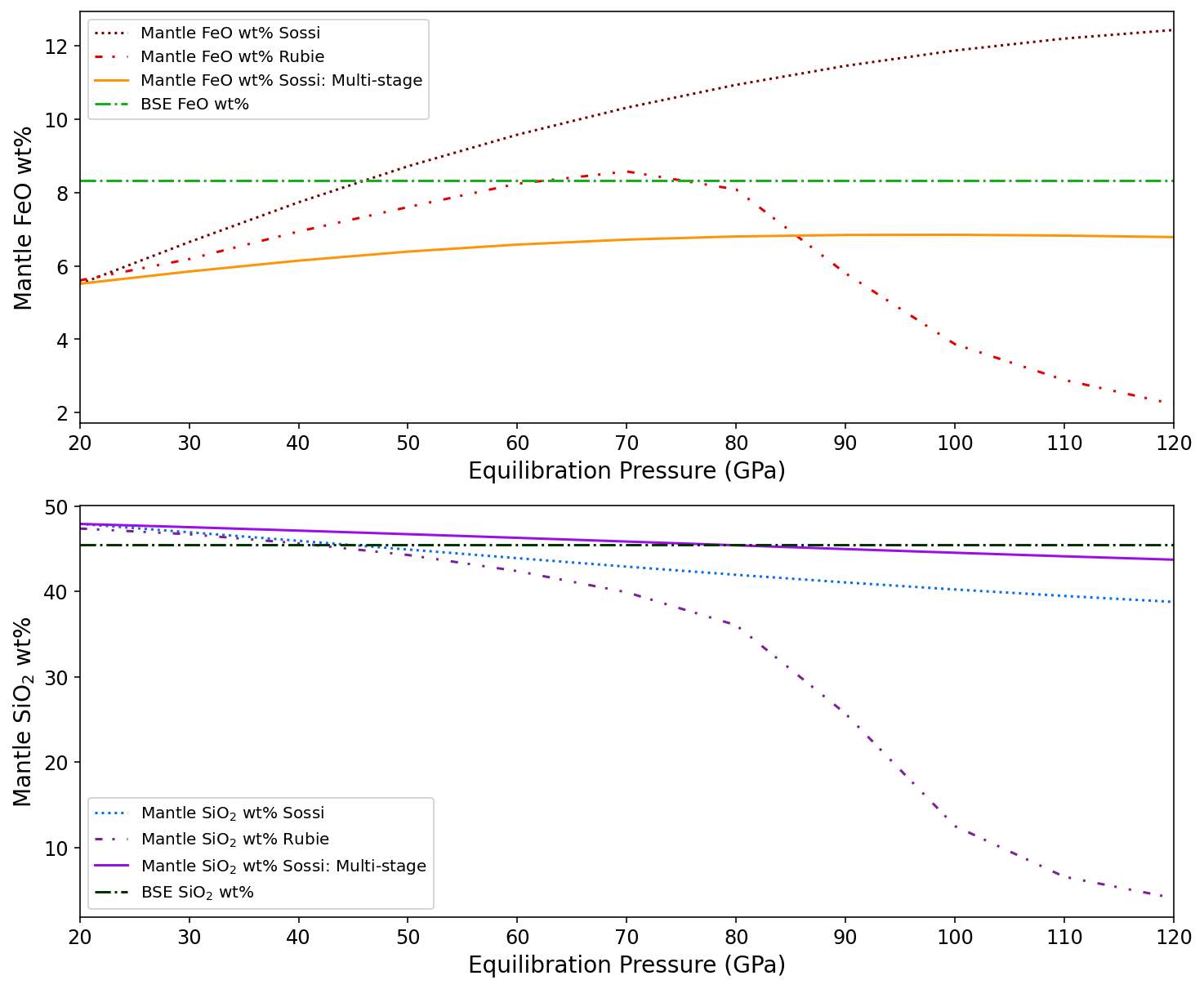}
    \caption{As \autoref{fig:PressurePlots} but additional lines (solid) showing the results of a simulated multi-stage accretion of the Earth using the partitioning parametrisations of \citet{Sossi2025}. These solid lines (in orange for FeO and purple for SiO$_2$) use the laws of \citet{Sossi2025} with 20\% (k$_{mantle}$ = 0.2) of the silicate interacting with the metal. The green dot-dashed lines on both figures show the BSE composition from \citet{Fischer2025}.}
    \label{fig:PressurePlots2}
\end{figure}

However, the situation changes radically if a multi-stage core formation scenario is considered. In a multi-stage scenario, Earth grows via a series of giant impacts from planetary embryos. In the case of homogeneous accretion, all embryos must have the bulk Earth composition otherwise the final Earth's bulk composition is not correct. The embryos initially differentiate due to the internal heating provided by radioactive decay, i.e following the single-stage scenario. Because of their small mass (0.1-0.35 Earth masses), this differentiation is expected to occur at about 20 GPa (and always less than 40 GPa), leading to FeO concentrations in their mantles smaller than that of the BSE (and to larger SiO$_2$ concentrations, see \autoref{fig:PressurePlots2}). Earth then grows via a sequence of giant impacts among these embryos. During these impacts, the metal of the core of the projectile equilibrates with only a small portion of the mantle of the target, typically of the order of 5-10\%, if the dispersal of iron droplets of the emulsified projectile’s core follows the description of \citet{Deguen2014}, or 20\% if it follows the prescription of \citet{Landeau2021}. The partial equilibration of the silicate with the metal becomes the key factor.

\begin{figure}[hbt!] 
    \centering
    \includegraphics[width=0.8\linewidth]{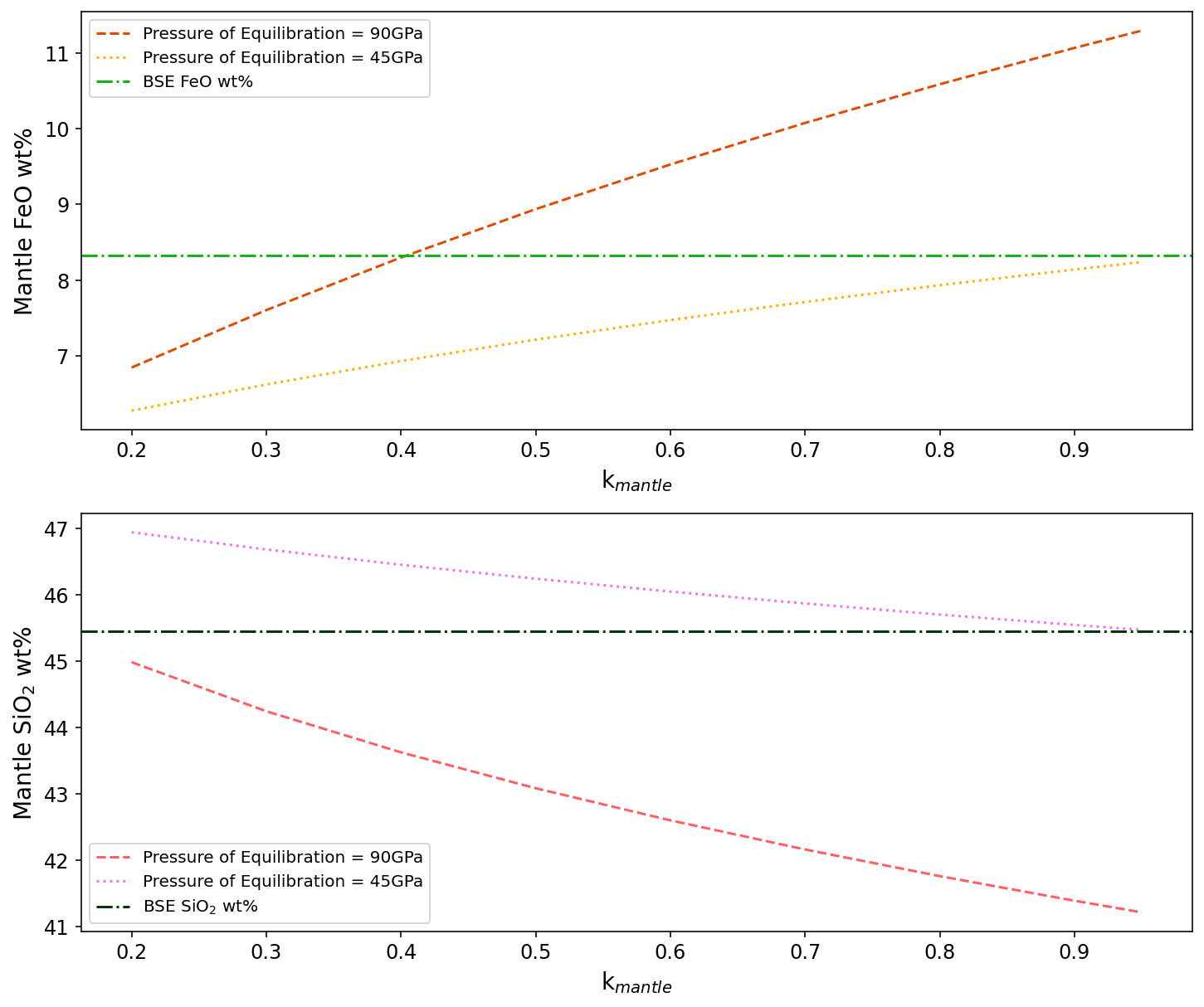}
    \caption{Final Earth mantle wt\% of FeO (top) and SiO$_2$ (bottom) if metal-silicate equilibration occurs at 45 GPa (dotted lines) and 90 GPa (dashed lines) as a function of k$_{mantle}$. k$_{mantle}$ is the fraction of the mantle that interacts with the metal during metal silicate equilibration, with k=1 representing full metal-silicate equilibration. For 90 GPa, no value of k$_{mantle}$ allows the BSE (green dot-dashed line) composition to be matched for both FeO and SiO$_2$. At 45 GPa, only a k$_{mantle}$ > 0.95 will match the BSE concentration of both oxides.}
    \label{fig:kmantle}
\end{figure}

To illustrate this point without using any specific accretion scenario, we consider an Earth that has initially the mantle and core compositions of the embryos (set via full metal-silicate equilibration at 20 GPa), and is then scrambled, leading to the reaction of all the core’s metal with a fraction k of the silicate at given pressure P.  Obviously if k=1 (full metal-silicate re-equilibration) and P=45GPa, using the partition law in \citet{Sossi2025} the BSE composition is reproduced. We then set k=0.2, the largest fraction of silicate reacting with the metal estimated for the giant impacts occurring in our dynamical model. The equilibration of the metal with this fraction of the silicate is computed at pressure P using the partition law of \citet{Sossi2025}. Then the equilibrated silicate is mixed with the 1-k fraction of the original silicate, which did not react with the metal. The orange and purple curves in \autoref{fig:PressurePlots2} show the resulting FeO and SiO$_2$ concentrations in the final, mixed mantle. The FeO never reaches the BSE value for any pressure. The equilibrated silicate has a high FeO fraction (12.8\% for P=90 GPa, for instance), but this high value is diluted when the equilibrated silicate is mixed with the silicate that did not react with the metal, leading to a too reduced Earth.

We also tested which value of k would be needed to reproduce the BSE in this scenario, for a given value of P, say 90 GPa, a value higher than the typical giant impact in our simulation. We find that no value of k allows us to reproduce simultaneously the FeO and SiO$_2$ fractions. For instance, SiO$_2$ is never reproduced while FeO requires k=0.4 (see \autoref{fig:kmantle}). Even if we reduce the pressure to 45 GPa (a still reasonable value for our simulations), which in single-stage core formation using the partitioning law of \citet{Sossi2025} recreates the BSE composition, k must be greater than 0.95 for both SiO$_2$ and FeO to be matched. From these tests, we conclude that the inability to reproduce the BSE chemistry in a homogenous multi-stage accretion scenario is due to the partial equilibration of the silicate with the iron characteristic of giant impacts and not to the quantitative properties of the partition law. 

Another observation argues against the reservoir of material of the terrestrial planets being homogeneous. The FeO concentration in the Martian mantle is $\sim$14\% \citep{Nathan2023}, much higher than that of the BSE. At no pressure, even in a single-stage differentiation scenario (which may be appropriate for Mars as a stranded planetary embryo), can one achieve such a high oxidation of the mantle starting from a bulk Earth composition \citep{Sossi2025}. Thus, the reservoir of material that led to the formation of terrestrial planets had to comprise of materials with varying oxidation states, which are accreted in different proportions by different growing planets. In this context, it is unlikely that Earth accreted the same material (i.e. homogenously) throughout time. 

\newpage
\section{Additional Figures and Tables}

\begin{figure} [hbt!] 
    \centering
    \includegraphics[width=0.65\linewidth]{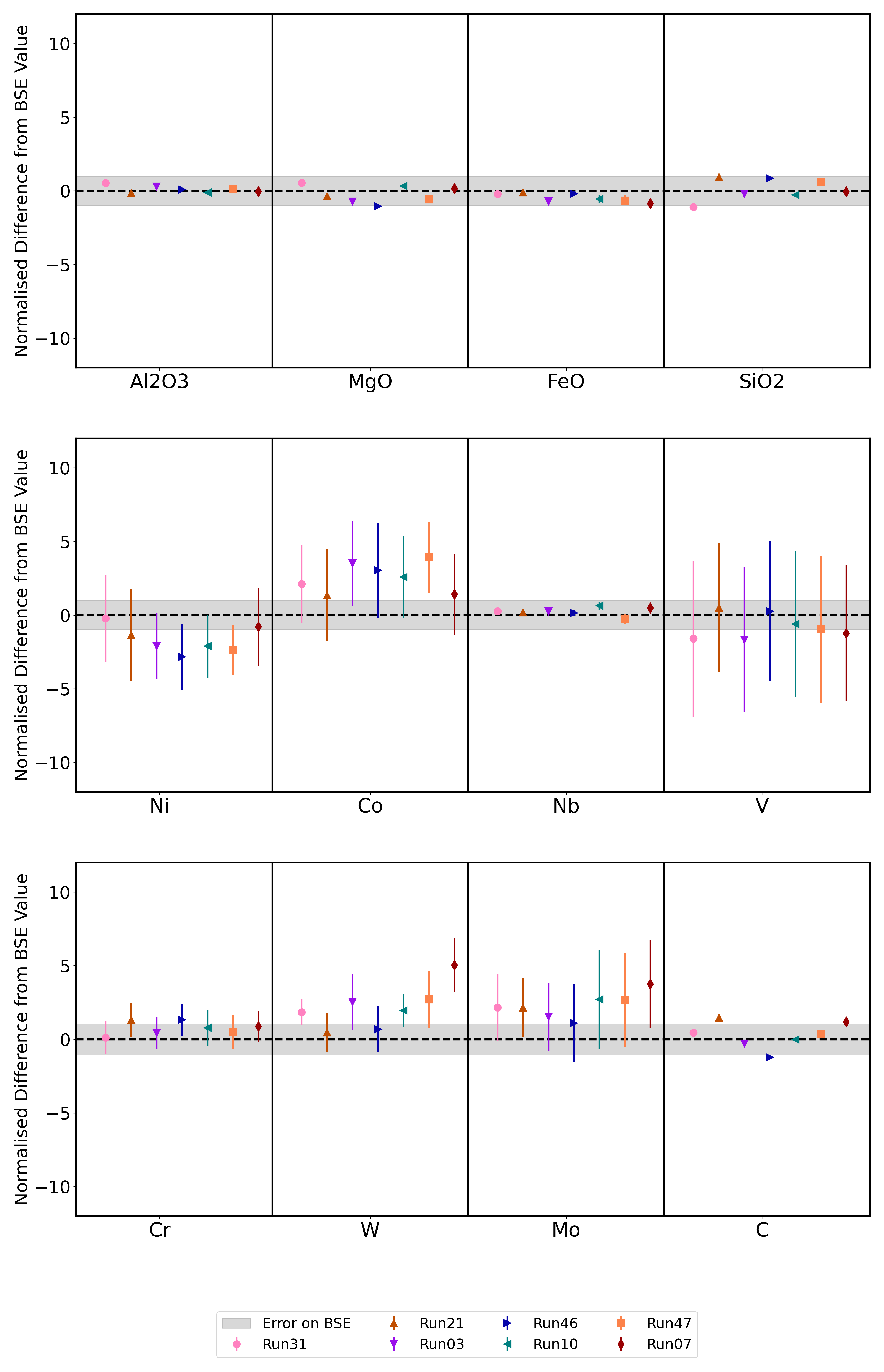}
    \caption{The composition of a selection of best Earth-analogues formed by the late addition of oxidised material to a reduced ring. Points are normalised as $C_{norm}^{model} = \frac{C^{model} - C^{BSE}}{\Delta C^{BSE}}$. We say a good fit for a given element is when the error bar for a given element (calculated solely from the propagation of errors on measurements of partition coefficients) crosses the error bar for BSE composition.}
    \label{fig:Sup4}
\end{figure}

\begin{figure}[hbt!] 
    \centering
    \includegraphics[width=0.7\linewidth]{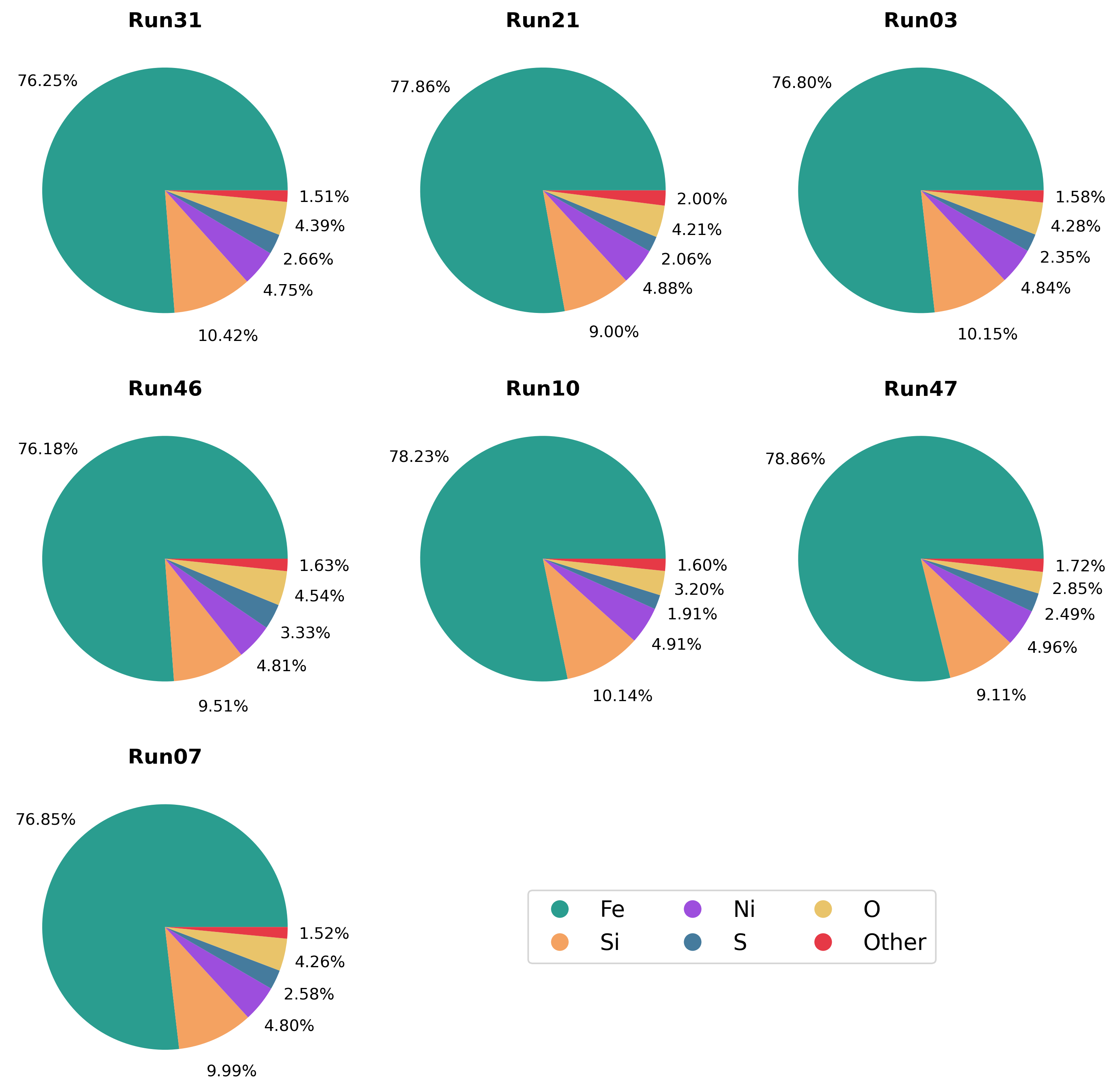}
    \caption{The core compositions of the 7 Earth analogues shown in \autoref{fig:Sup1}. Percentages are wt\% of the core, core mass fractions can be found in \autoref{fig:Sup1}.}
    \label{fig:Sup2}
\end{figure}
 
\begin{table*}[hbt!]
  \centering
  \caption{From \citet{Dale2025}. The compositions of bodies in each distinct region of the ring with meteoritic data from \cite{Alexander2019, Alexander2019b}. Carbon contents are assigned following \cite{Blanchard2022} with undifferentiated NC bodies containing 80 times more carbon than their differentiated counterparts. Undifferentiated CC bodies contain 1500 times more carbon than differentiated CCs. In the table, all bodies are embryos and thus differentiated, though CCs contain no metal and have lost carbon due to heating. Refractory enriched bodies are entirely carbon-depleted. Water contents for CIs are taken from (\cite{Alexander2019}) while OCs contain a small quantity of water (e.g. \citealp{Hutchison1987,Piani2015,Grant2023}). Quantities marked with a * are enriched relative to CI ratios by a factor of 1.3 in this example but this parameter is free to vary through the model. Mantle oxides are listed in the first half of the table and are highlighted in grey. Core metals are in the lower half. Initial partitioning occurs at 0.65 x core mantle boundary pressure and the associated temperature, though we find this makes little difference to the initial compositions of the bodies. BSE oxide abundances are from \citet{Palme&ONeill2003} with the exception of Mo, which is from \citet{Greber2015}}.
  \label{tab:Compostions}
  \renewcommand{\arraystretch}{0.9} 

    \begin{tabular}{| p{1.5cm} | p{1.8cm} | p{1.8cm} | p{1.8cm} | p{1.8cm} | p{3.15cm} |}
    \hline
    & \multicolumn{4}{|c|}{Zone Compositions (wt\%)} & \\ \hline
Component & Refractory \newline Enriched & Enstatite Chondrite & Ordinary Chondrite & CI Chondrite & BSE \\ \hline
\rowcolor{LightGrey}
    Al  &  3.548228*  & 1.897557  & 2.170483   & 1.710768 & 4.49 $\pm$ 0.359 \\ \hline
\rowcolor{LightGrey}
    Mg   & 26.201399   & 22.481138    & 23.782349   & 17.054456  & 36.77 $\pm$ 0.368\\ \hline
\rowcolor{LightGrey}
    Ca  & 2.849528*  & 1.481915  & 1.770680  & 1.373892 & 3.65 $\pm$ 0.292 \\ \hline
\rowcolor{LightGrey}
    Na  & 0.000  & 0.935748  & 0.892294  & 0.720640  & 0.259 $\pm$ 0.013\\ \hline
\rowcolor{LightGrey}
    Fe  & 0.000  & 0.000  & 21.133129  & 25.652474 & 8.1 $\pm$ 0.050 \\ \hline
\rowcolor{LightGrey}
    Si  & 31.459577  & 39.881963  & 37.691678  & 24.671080 & 45.4 $\pm$ 0.318 \\ \hline
\rowcolor{LightGrey}
    Ni  & 0.022336  & 0.021299  & 0.015769  & 1.453873 & 0.186 $\pm$ $9.3 \times 10^{-3}$ \\ \hline
\rowcolor{LightGrey}
    Co  & 0.001093  & 0.001054  & 0.000780  & 0.071128  & 0.0102 $\pm$ $5.1 \times 10^{-4}$\\ \hline
\rowcolor{LightGrey}
    Nb  & $1 \times 10^{-6}$* & 0.000  & $1 \times 10^{-6}$  & $4.4 \times 10^{-5}$  & $5.95\times10^{-5}$ $\pm$ $1.19\times10^{-5}$ \\ \hline
\rowcolor{LightGrey}
    V & $1.8 \times 10^{-4}$*  & $9 \times 10^{-5}$  & $1.08 \times 10^{-4}$  & $8.658 \times 10^{-3}$  & $8.6\times10^{-3}$ $\pm$ $4.3\times10^{-4}$\\ \hline
\rowcolor{LightGrey}
    Cr  & $5.673 \times 10^{-3}$   & $4.543 \times 10^{-3}$  & $4.786 \times 10^{-3}$  & 0.369284  & 0.252 $\pm$ 0.0252\\ \hline
\rowcolor{LightGrey}
    Pt  & $2 \times 10^{-6}$*  & $1 \times 10^{-6}$  & $1 \times 10^{-6}$  & $1.01 \times 10^{-4}$  & $7.6\times10^{-7}$ $\pm$ $1.52\times10^{-7}$\\ \hline
\rowcolor{LightGrey}
    Pd  & $1 \times 10^{-6}$ & $1 \times 10^{-6}$  &$1 \times 10^{-6}$  & $6.9 \times 10^{-5}$ & $7.1\times10^{-7}$ $\pm$ $1.42\times10^{-7}$ \\ \hline
\rowcolor{LightGrey}
    Ru & $2 \times 10^{-6}$* & $1 \times 10^{-6}$  &$1 \times 10^{-6}$  & $6.9 \times 10^{-5}$ & $7.4\times10^{-7}$ $\pm$ $1.48\times10^{-7}$ \\ \hline
\rowcolor{LightGrey}
    Ir & $1 \times 10^{-6}$* & $1 \times 10^{-6}$  &$1 \times 10^{-6}$  & $5.4 \times 10^{-5}$ & $3.5\times10^{-7}$ $\pm$ $3.5\times10^{-8}$ \\ \hline
\rowcolor{LightGrey}
    W & 0.000* & 0.000  & 0.000 & $1.3 \times 10^{-5}$  & $1.2\times10^{-6}$ $\pm$ $5.04\times10^{-7}$\\ \hline
\rowcolor{LightGrey}
    Mo & $3 \times 10^{-6}$* & $2 \times 10^{-6}$  &$1 \times 10^{-6}$  & $1.38 \times 10^{-4}$ & $2.3\times10^{-6}$ $\pm$ $1.38\times10^{-6}$  \\ \hline
\rowcolor{LightGrey}
    H$_2$O & 0.000 & 0.000  & 0.973456 & 21.556782 & 0.1 $\pm$ 0.03\\ \hline
\rowcolor{LightGrey}
    S & 0.000 & 0.000  & 0.000 & 11.520290 & 0.02 $\pm$ $5.0 \times10^{-3}$ \\ \hline
\rowcolor{LightGrey}
    C & 0.000 & 0.000  & 0.000 & 0.005027  & 0.014 $\pm$ $4.0 \times10^{-3}$  \\ \hline

    Fe & 30.634604 & 27.986974  & 5.475691 & 0.000 & \\ \hline

    Si & 3.012127 & 0.837689  & 0.000 & 0.000 &\\ \hline

    Ni & 1.737727 & 1.656994  & 1.226818 & 0.000 & \\ \hline

    Co & 0.085083 & 0.082035  & 0.060714 & 0.000 & \\ \hline

    Nb & $6.3 \times 10^{-5}$* & $2.9 \times 10^{-5}$  &$3.6 \times 10^{-5}$  & 0.000  & \\ \hline

    V & 0.012084* & 0.006030  & 0.007257  & 0.000  & \\ \hline

    Cr & 0.429514 & 0.343894  & 0.362359  & 0.000 & \\ \hline

    Pt & $1.91 \times 10^{-4}$* & $1.34\times 10^{-4}$  &$1.12 \times 10^{-4}$  & 0.000  &\\ \hline

    Pd & $9.2 \times 10^{-5}$ & $8.8 \times 10^{-5}$  &$6.5 \times 10^{-5}$  & 0.000 & \\ \hline

    Ru & $1.53 \times 10^{-4}$* & $9.4 \times 10^{-5}$  &$7.9 \times 10^{-5}$  & 0.000 &\\ \hline

    Ir & $1.02 \times 10^{-4}$* & $5.6 \times 10^{-5}$  &$5.1 \times 10^{-5}$  & 0.000 & \\ \hline

    W & $2.1 \times 10^{-5}$* & $1.5 \times 10^{-5}$  &$1.4 \times 10^{-5}$  & 0.000  &\\ \hline

    Mo & $2.13 \times 10^{-4}$* & $1.15 \times 10^{-4}$  & $1.11 \times 10^{-4}$  & 0.000 & \\ \hline

    S & 0.000 & 2.378344  & 4.429226  & 0.000  &\\ \hline

    C & 0.000 & 0.002195  & 0.001947  & 0.000 & \\ \hline
\end{tabular}
\end{table*}

\begin{table*}[hbt!]
    \centering
  \caption{The oxidation of each distinct region of the ring. Iron oxidation is taken from the Urey-Craig diagram (e.g. \citealp{Krot2014}). Silicon oxidation in enstatite chondrites is calculated using \citet{Keil1968} and \citet{Lehner2014}. Total oxygen in each body at the beginning of a simulation is calculated using this and assigning lithophile elements as 100\% oxidised and siderophile elements as 1\% oxidised (see \autoref{tab:Compostions} for 'siderophile' and 'lithophile' elements).}
  \label{tab:Oxidation}
  \begin{tabular}{|c|c|c|c|c|}
    \hline
    & \multicolumn{4}{|c|}{Zone Oxidation (\%)} \\ \hline
     & Refractory Enriched & Enstatite Chondrite & Ordinary Chondrite & CI Chondrite \\ \hline
     Percentage Si in Metal & 17  & 4.3  & 0.0  & 0.0 \\ \hline
     Percentage Fe in Metal & 100  & 100  & 25  & 0.0 \\ \hline
    \end{tabular}
\end{table*}

\end{document}